\documentclass[pra]{revtex4}

\usepackage{amsmath, amssymb, amsthm, amsfonts, latexsym, graphicx, mathdots, color}

\newcommand{\half}{\frac{1}{2}}

\newcommand{\ket}[1]{\left| #1\right\rangle}      
\newcommand{\kets}[1]{| #1 \rangle}                    
\newcommand{\bras}[1]{\langle #1 |}                   
\newcommand{\braket}[2]{\langle #1 | #2 \rangle}   
\newcommand{\norm}[1]{\left\| #1\right\|}          
\newcommand{\ep}{\epsilon}                               

\begin{document}

\title{Quantum Walks on Necklaces and Mixing}

\author{M\'aria Kieferov\'a}
\affiliation{Research Center for Quantum Information, Institute of Physics, Slovak Academy of Sciences,
D\'{u}bravsk\'{a} cesta 9, 845 11 Bratislava, Slovakia}
\author{Daniel Nagaj}
\affiliation{Research Center for Quantum Information, Institute of Physics, Slovak Academy of Sciences,
D\'{u}bravsk\'{a} cesta 9, 845 11 Bratislava, Slovakia}
\email{daniel.nagaj@savba.sk}

\begin{abstract}
We analyze continuous-time quantum walks on necklace graphs
-- cyclical graphs consisting of many copies of a smaller graph (pearl). 
Using a Bloch-type ansatz for the eigenfunctions, we block-diagonalize the Hamiltonian, reducing
the effective size of the problem to the size of a single pearl. 
We then present a general approach for showing that the mixing time scales
(with growing size of the necklace) similarly to that of a simple walk on a cycle. 
Finally, we present results for mixing on several necklace graphs. 
\end{abstract}

\date{\today}


\maketitle


\section{Introduction}
\label{introduction}

Classical random walks on form the basis for many successful physics-inspired algorithms.
The evolution of probability distributions according to simple update rules for probability spreading
allows us to sample from thermal distributions (via the Metropolis algorithm \cite{metropolis, metropolis2}) or to look for ground-states of physical systems(with simulated annealing \cite{annealing, annealing2}). 
The effectiveness of random-walk based algorithms can be characterized by its mixing time (how fast it approaches the stationary distribution), or by a hitting time (how fast it reaches a particular vertex). For example, the fast mixing of a random walk algorithm for sampling from the thermal distribution of the Ising model \cite{Jerrum2} forms the basis of a fully polynomial randomized approximation scheme for the permanent of a matrix \cite{Vigoda}.

Thinking about how to utilize the probabilistic nature of quantum mechanics, instead of analyzing the diffusion of probabilities, we can ask what happens if we let the amplitudes in a system whose interactions respect some graph structure evolve according to the
Schr\"{o}dinger equation. The result of this way of thought are {\em quantum walks} \cite{AAKVwalk, FarhiWalk}, a useful tool in quantum computation. They bring new dynamics (different wavepacket spreading \cite{ABNVWwalk}) and algorithmic applications (e.g. in searching for graph properties \cite{Santha}, graph traversal
\cite{gluedtrees}, game evaluation 
\cite{nandtrees}) as well as theoretical results (universality for computation \cite{universal}). We can define quantum walks in
discrete time with an additional coin register, or in continuous time, with Hamiltonians which are adjacency matrices of graphs. In this paper,
we choose the latter approach.

The mixing of quantum walks has been previously investigated for several types of graphs -- e.g. on a chain \cite{AAKVwalk}, a 2D lattice \cite{2Dmix}, hypercubes \cite{hypercubemix} and circulant graphs \cite{circulant1, circulant2}. 
In this paper we focus on continuous quantum walks on {\em necklaces} -- cyclic graphs composed from many ($K$) copies of a subgraph of size $M$ (pearls), as depicted in Fig.\ref{figurenecklaces}. Our goal is to provide a simplified approach for finding their eigenvectors and eigenvalues, as well as for analyzing the mixing times for such walks.  

The motivation for analyzing this type of graph comes from Hamiltonian complexity \cite{HamiltonComplex}. Quantum computation in the usual circuit model \cite{NCbook} can be translated into a quantum walk in two ways. First, following Childs
\cite{universal}, evolving a wavepacket on a graph with many wires (representing basis states), connected according to the desired quantum circuit.
Second, we can use Feynman's idea \cite{Feynman85} to view a computation as a ``pointer'' particle doing a quantum walk (hopping) in a ``clock''
register, while the computation gets done in a ``data'' register \cite{deFalco, KKR, new3local, railroadswitch} or particles holding the working data hopping along a graph \cite{qma1d, LidarMizel}. In both cases, we need to
look at transmission/reflection properties of the graphs, and their long-term dynamics. Specifically, we would like to know (and ensure) that a
computation is done when we want it to be, not having the wavepacket localized (or spread) in undesired parts of the graph. This is why we focus on the
mixing properties of quantum walks that are underlying quantum computational models based on quantum walks, looking at their spectra in detail. Note that proofs of computational complexity for QMA-hard problems (e.g. \cite{KKR, new3local}) also involves investigating the (low-lying) spectrum of a quantum walk. The simplest graph involved in the Feynman-like models is a line or a cycle, and the dynamics for this quantum walk are well understood
\cite{AAKVwalk, deFalco}. We look at continuous-time quantum walks on {\em necklace} graphs, which appear in the analysis of quantum computational models \cite{railroadswitch, railroadswitch2, EldarRegev} that generalize the Feynman approach. 
Necklace graphs could also be viewed as implementing dynamics for quantum walks on imperfect cycles.

Utilizing the cyclic structure of the necklaces, we propose a Bloch-type ansatz for the eigenfunctions, allowing us to obtain several results.
First, in Section \ref{sec:ansatz} we reduce the problem of finding the eigenvectors and eigenvalues of the quantum walk on a necklace of $K$
pearls of size $M$ to diagonalizing a $M\times M$ matrix $K$ times (compared to full $KM\times KM$ diagonalization. Second, in Section
\ref{sec:mixing}, we analyze average-time mixing for quantum walks on necklaces and find a general method for showing convergence to the limiting
distribution. Finally, in Section \ref{sec:examples}, we work out examples of quantum walks on particular necklaces, giving analytic (and
numerical) results for the eigenvectors, eigenvalues and the scaling of the mixing time, concluding with open questions in
Section \ref{sec:conclusions}. 


\section{Finding eigenvectors and eigenvalues}
\label{sec:ansatz}

\begin{figure}
   \begin{center}
      \includegraphics[width=3in]{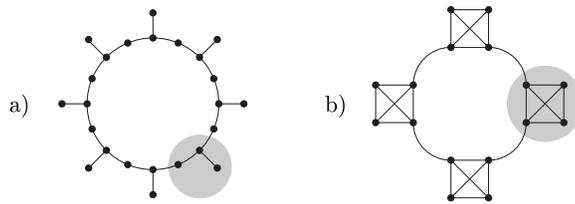}
   \end{center}
   \caption{Examples of necklace graphs, with the pearls denoted by shaded regions.}
   \label{figurenecklaces}
\end{figure}

Consider a quantum system with a Hamiltonian $H$ given by the adjacency matrix of a 
necklace-like structure. The simplest necklace is a cycle with $K$ vertices.
A~general necklace is a collection of $K$ pearls (small identical graphs with $M$ nodes),
connected into a cycle as in Fig. \ref{figurenecklaces}.

We label points in the $j$-th pearl $x^{(j)}_{m}$, with $1\leq m \leq M$. The endpoints of the $j$-th pearl (connected to the previous and
following pearls) are $x_1^{(j)}$ and $x_M^{(j)}$ Let $P$ be the adjacency matrix of a pearl.
The Hamiltonian for the whole necklace is a sum of intra-pearl terms and the connections between them: 
\begin{eqnarray}
   H = \sum_{j=1}^K P^{(k)}  
      &+& \sum_{j=1}^{K-1} \left( \kets{x^{(j)}_M}\bras{x^{(j+1)}_1}
                                          + \kets{x^{(j+1)}_1}\bras{x^{(j)}_M} \right) \label{Hsum} \\
      &+& \left( \kets{x^{(K)}_M}\bras{x^{(1)}_1}
                                          + \kets{x^{(1)}_1}\bras{x^{(K)}_M} \right). \nonumber
\end{eqnarray}

Our goal is to find the eigenvalues and eigenvectors of $H$. 
Because of the underlying cyclic structure of a general necklace graph with $K$ pearls,
we can assume that its eigenvectors will have a structure related to a plane wave on a cycle with $K$ nodes. 
Let us then look at the $K$-node cycle first. There the Hamiltonian \eqref{Hsum} has no $P^{(k)}$'s in it, 
allowing us to find the (plane-wave) eigenvectors of $H^{\circ}$:
\begin{eqnarray}
   \ket{w_k^\circ} = \sum_{j=1}^{K} e^{i p_k j} \ket{x_j}, \label{cyclevectors}
\end{eqnarray}
corresponding to eigenvalues $\lambda_k$ parametrized by momenta $p_k$:
\begin{eqnarray}
   \lambda_k^\circ = 2 \cos p_k^\circ, \qquad 
   p_k^\circ = \frac{2\pi k}{K},
   \label{momentum}
\end{eqnarray}
for $k=0,\dots,K-1$.

\begin{figure}
   \begin{center}
      \includegraphics[width=2.6in]{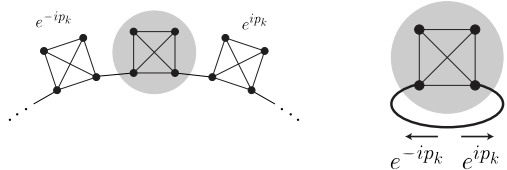}
   \end{center}
   \caption{Using the plane wave-like ansatz \eqref{ceigenfunctions} in which neighboring pearls get a constant phase factor difference, the
necklace Hamiltonian can be block-diagonalized on the pearls, acting independently on each pearl with the addition of a single link (carrying a
phase factor) between its roots $x_1$ and $x_M$.}
   \label{figurenecklacePB}
\end{figure}
Consider now a general necklace with $K$ pearls. We expect the eigenvectors of the necklace Hamiltonian \eqref{Hsum}
to have a form resembling \eqref{cyclevectors}, also depending on the momenta $p_k$ \eqref{momentum}. Let us thus look
for the eigenvectors of $H$ in the form
\begin{eqnarray}
   \ket{\psi_k} 
   = \frac{1}{\sqrt{K}} \sum_{j=1}^{K} e^{i p_k j} \kets{y_k^{(j)}},
   \label{ceigenfunctions}
\end{eqnarray}
where each
\begin{eqnarray}
   \kets{y_k^{(j)}} = \sum_{m=1}^{M} y^{k}_{m} \kets{x_{m}^{(j)}}
   \label{whatisy}
\end{eqnarray}
is a normalized vector with support only on the $j$-th pearl (the vertices $x_{1}^{(j)}, \dots, x_{M}^{(j)}$).
Using \eqref{Hsum} and \eqref{ceigenfunctions}, we obtain
\begin{eqnarray}
   H \ket{\psi_k} &=& \frac{1}{\sqrt{K}} \sum_{j=1}^{K} e^{i p_k j} 
   \left(
         P^{(j)} \kets{y_k^{(j)}}
      + e^{-i p_k} y^{k}_{M} \kets{x^{(j)}_1}
      + e^{i p_k} y^{k}_{1} \kets{x^{(j)}_M}
   \right)
   \label{usedansatz}
\end{eqnarray}
where the last two terms correspond to the amplitudes on the endpoints of the $j$-th pearl
coming from the endpoints of the neighboring pearls. Notice that because of our parametrization \eqref{ceigenfunctions},
the Hamiltonian is now block-diagonalized, acting in the same way on each pearl (see Figure \ref{figurenecklacePB}).
When $\ket{\psi_k}$ is an  eigenvector of $H$, we also have
\begin{eqnarray}
   H \ket{\psi_k} &=& \lambda_k \ket{\psi_k}
      = \lambda_k \frac{1}{\sqrt{K}} \sum_{j=1}^{K} e^{i p_k j} \kets{y_k^{(j)}}.
      \label{necklaceeigenvector}
\end{eqnarray}
Using \eqref{usedansatz} and \eqref{necklaceeigenvector}, finding the eigenvalues of $H$ thus reduces to diagonalizing
the $M\times M$ matrix
\begin{eqnarray}
   Y_k = P + Q_k, \label{whattodiag}
\end{eqnarray}
where $P$ is the adjacency matrix of a pearl, and  
\begin{eqnarray}
   Q_k = \left[\begin{array}{ccccc}
                  && &  & e^{-ip_k} \\
                   &  &&0& \\
                   &  & \iddots && \\
                   & 0 &&& \\
                  e^{i p_k} &  & && 
               \end{array}
            \right]
\end{eqnarray}
has only two non-zero elements in the corners if a pearl has two distinct roots $x_{1}$ and $x_{M}$. 
There is a special case when a pearl is connected to the rest of the necklace through a single root vertex $x_{1}$. There,  
the matrix $Q_k$ has a single nonzero element and reads
\begin{eqnarray}
   Q_k = \left[\begin{array}{ccc}
                   e^{ip_k} + e^{-ip_k} &  &\\
                   & 0 & \\
                   & & \ddots
               \end{array}
            \right].
\end{eqnarray}
Diagonalizing \eqref{whattodiag} gives us $M$-dimensional vectors $\kets{y_k}$.
For each $k = 0,\dots,K-1$, there will be $M$ of these, and we will label them $\kets{y_{k,n}}$ with $n=1,\dots M$.
The corresponding eigenvalues $\lambda_{k,n}$ of $Y_k$ are also the eigenvalues of the full Hamiltonian $H$.
Therefore, to find all the $KM$ eigenvalues $\lambda_{k,n}$ of the necklace Hamiltonian with $K$ pearls, we need to
diagonalize the $M\times M$ matrix $Y_k$ \eqref{whattodiag} for each $k = 0,\dots,K-1$. 
To get the eigenvectors of $H$ from the eigenvectors of $Y_k$,
we plug the coefficients $y^{(k,n)}_m$ of the vectors we just found into \eqref{ceigenfunctions} and \eqref{whatisy}.

In conclusion, the ansatz \eqref{ceigenfunctions} simplifies the general problem of diagonalizing the $KM\times KM$
matrix $H$ 
to diagonalizing an $M\times M$ matrix $K$ times.
This is useful especially when $M$ is small and $K$ is large. Our focus in what follows will be on mixing of continuous quantum walks on
many-pearled (large-$K$) necklaces.


\section{Quantum walks and mixing}
\label{sec:mixing}

\subsection{Mixing in a time-averaged sense}
Time evolution according to the Schr\"{o}dinger equation with a Hamiltonian that is an adjacency
matrix of a graph produces a continuous time quantum walk.
Let the eigenvectors of the system be $\ket{\psi_{k}}$ and the corresponding
eigenvalues $\lambda_k$. When starting from an initial state $\ket{\varphi_0}$,
the probability of finding the ``walker'' at vertex $\ket{x}$ at time $t$ (measuring position $x$) is
\begin{eqnarray}
   p_{x}^{\ket{\varphi_0}}(t) 
   &=& \left|\braket{x}{\varphi(t)}\right|^2 
   = \left|\sum_{k} e^{-i\lambda_k t}\braket{x}{\psi_k}\braket{\psi_k}{\varphi_0}\right|^2.
\end{eqnarray}
The evolution is unitary, so it does not mix towards a time-independent stationary distribution
like a classical Markov process. On the other hand, we can think about mixing for a quantum walk 
in a time-averaged sense, investigating a time-averaged probability distribution. It holds information about the probability of finding the system at a particular
vertex at time $t$, chosen uniformly at random between $0$ and $T$ (a chosen limiting time):
\begin{eqnarray}
   \bar{p}_{x}^{\ket{\varphi_0}}(T) 
   &=& \frac{1}{T}\int_{\tau=0}^{T} p_{x}^{\ket{\varphi_0}}(\tau).
\end{eqnarray}
This time-averaged probability has a well-defined $T\rightarrow \infty$ limit, which gives us the
{\em limiting probability distribution}, expressible using the eigenvectors of $H$ as:
\begin{eqnarray}
   \pi_x^{\varphi_0} &=& \lim_{T\rightarrow \infty} \bar{p}_{x}^{\ket{\varphi_0}}(T) \label{limprob}\\ 
   &=& \lim_{T\rightarrow \infty} \frac{1}{T} \int_{\tau=0}^{T} 
      \left(\sum_{k} e^{-i\lambda_k \tau}\braket{x}{\psi_k}\braket{\psi_k}{\varphi_0}\right)
      \left(\sum_{l} e^{i\lambda_l \tau}\braket{\psi_l}{x}\braket{\varphi_0}{\psi_l}\right)\,d\tau \\
   &=& \sum_{k}\sum_{l} 
   \braket{x}{\psi_k}\braket{\psi_k}{\varphi_0}
   \braket{\psi_l}{x}\braket{\varphi_0}{\psi_l}
   \left(\lim_{T\rightarrow \infty} \frac{1}{T} \int_{\tau=0}^{T} e^{-i (\lambda_k-\lambda_l) \tau}\,d\tau\right) 
   \\
   &=& \sum_{\lambda_k=\lambda_l}
   \braket{\psi_l}{x}\braket{x}{\psi_k}
   \braket{\psi_k}{\varphi_0}
   \braket{\varphi_0}{\psi_l}, \label{limiting}
\end{eqnarray}
where the final sum goes over pairs of equal eigenvalues.
Note that for some quantum walks this limiting distribution can be dependent on the initial state 
(e.g. when we start in some eigenstate), so we will keep the superscript $\varphi_0$ around.

To determine how fast the time-averaged probability converges towards the limiting distribution,
we need to bound the total distribution distance $\norm{\bar{p}^{\varphi_0}(T)-\pi^{\varphi_0}}$.
Using \eqref{limiting}, integrating an exponential and realizing that the terms
summed over pairs of equal eigenvalues subtract out, we arrive at
\begin{eqnarray}
   \norm{\bar{p}^{\varphi_0}(T)-\pi^{\varphi_0}} 
      = \sum_{x} \left| \bar{p}^{\varphi_0}_x - \pi_x^{\varphi_0} \right| 
      &=& \sum_{x} \left| \sum_{\lambda_k \neq \lambda_l}
      \braket{\psi_l}{x}\braket{x}{\psi_k}
      \braket{\psi_k}{\varphi_0}
      \braket{\varphi_0}{\psi_l}
      \left(
      \frac{e^{-i(\lambda_k-\lambda_l)T}-1}{-i(\lambda_k-\lambda_l)T}
      \right)
      \right|.
\end{eqnarray}
where the sum now goes over pairs of eigenvalues that are not equal.
We can put an upper bound on this expression by a technique similar to \cite{AharonovWalks}.
First, we use $|e^{-i(\lambda_k-\lambda_l)T}-1|\leq 2$ and move the absolute value inside the sums, to obtain
\begin{eqnarray}
   \norm{\bar{p}^{\varphi_0}(T)-\pi^{\varphi_0}} 
      \leq 
      \sum_{x} \sum_{\lambda_k \neq \lambda_l}
      \left| 
      \braket{\psi_l}{x}
      \right|
      \left| 
      \braket{x}{\psi_k}
      \right|
      \frac{2
         \left| 
         \braket{\psi_k}{\varphi_0}
         \right|
         \left| 
         \braket{\varphi_0}{\psi_l}
         \right|
      }{T|\lambda_k-\lambda_l|}.
\end{eqnarray}
The Cauchy-Schwartz inequality $\left|\braket{\psi_l}{x}\right|\left|\braket{x}{\psi_k}\right|
\leq \half \left(
\left|\braket{\psi_l}{x}\right|^2 + \left|\braket{x}{\psi_k}\right|^2
\right)$
allows us to perform the sum over $x$, resulting in 
\begin{eqnarray}
   \norm{\bar{p}^{\varphi_0}(T)-\pi^{\varphi_0}}
      \leq 
      \sum_{\lambda_k \neq \lambda_l}
      \frac{2
         \left| 
         \braket{\psi_k}{\varphi_0}
         \right|
         \left| 
         \braket{\varphi_0}{\psi_l}
         \right|
      }{T|\lambda_k-\lambda_l|}.
\end{eqnarray}
After another use of the Cauchy-Schwartz inequality on the terms involving $\ket{\varphi_0}$, realizing the expression
is symmetric under exchange of $k$ and $l$, we finally obtain 
\begin{eqnarray}
  \norm{\bar{p}^{\varphi_0}(T)-\pi^{\varphi_0}}
      \leq 
      \sum_{\lambda_k \neq \lambda_l}
      \frac{2
         \left| 
         \braket{\psi_k}{\varphi_0}
         \right|^2
      }{T|\lambda_k-\lambda_l|},
      \label{Lemma43}
\end{eqnarray}
which corresponds to Lemma 4.3 of \cite{AharonovWalks}. It involves a sum of the inverse of eigenvalue differences.
These terms can be large, but as $T$ grows, the $1/T$ factor can bring the total variation difference to zero. It is our
task now to investigate how fast this happens. We seek $T_{mix}$ (the {\em mixing
time}), for which
\begin{eqnarray}
      \norm{\bar{p}^{\varphi_0}(T)-\pi^{\varphi_0}} \leq \ep
      \label{mixingtime}
\end{eqnarray}
would hold for all $T\geq T_{mix}\left(\ep\right)$, given any precision parameter $\ep$.


\subsection{Quantum Walk on a Cycle: The Limiting Distribution}
For our first example, we now follow \cite{HQCA1d} and compute the limiting distribution for the case of a walk on a cycle.
Later, we will show that the time-averaged probability converges to it for times $T = O(K \log K)$,
using a more general mixing result proved in Section \ref{sec:method}.

The eigenvalues and eigenvectors for the continuous-time quantum walk on a cycle are given by \eqref{cyclevectors} and \eqref{momentum}. We obtain
the limiting distribution from 
\eqref{limiting} by summing over the few nonzero terms.
The sum over the equal eigenvalues splits into a sum over $k=l$ and $k+l=K$ (degenerate eigenvalue pairs). 
When the initial state $\ket{\varphi_0}$ is concentrated at a vertex $z$, 
in the case of even $K$, the limiting distribution 
for the quantum walk on a cycle is
\begin{eqnarray}
   \pi(x|z) &=& \frac{1+f_{x,z}}{K} - \frac{2}{K^2}, \label{fxz}
\end{eqnarray}
where $f_{x,z}=0$ for all pairs $(x,z)$, with an exception for the two points $x^*=z$ and $|x^*-z|=\frac{K}{2}$, where its value is $f_{x^*,z}=1$.
For a cycle with an odd length $K$, we get
\begin{eqnarray}
   \pi(x|z) &=& \frac{1+f_{x,z}}{K} - \frac{1}{K^2},
\end{eqnarray}
with $f_{x,z}$ defined in the same way as for even $K$, equal to zero for all pairs $(x,z)$ except for $x^*=z$, where $f_{z,z}=1$.
The slight differences from a uniform distribution arise because not all of the eigenvalues are doubly degenerate.

Proving that the time-averaged distribution converges towards the limiting distribution
for $T \geq O(K \log K)$ takes more work.
We want to show that the total distribution distance $\norm{\bar{p}^{\varphi_0}(T)-\pi^{\varphi_0}}$ goes to zero as $T \geq O(K \log K)$.
When computing $\bar{p}^{\varphi_0}(T)$, the terms with $\lambda_k=\lambda_l$ produce the limiting distribution $\pi^{\varphi_0}$ and are thus
subtracted out.
However, the terms left over (which were killed by the $T\rightarrow\infty$ limit when computing $\pi$) need to be carefully accounted for and
govern the convergence.
In \eqref{Lemma43}, we have a bound on the total distribution distance
by a sum over pairs of inequal eigenvalues.
We will upper bound this sum in Section \ref{sec:method}, using a general approach of lower bounding the terms $|\lambda_k-\lambda_l|^{-1}$ in
\eqref{Lemma43}. This result is then applicable to several other walks on necklaces.


\subsection{A general approach to proofs of mixing}
\label{sec:method}

The rate of convergence of the time-averaged distribution towards the limiting distribution is governed by a sum of
$|\lambda_{k,n}-\lambda_{j,m}|^{-1}$ over non-equal eigenvalues as in \eqref{Lemma43}. We will now show a method for upper bounding it that will
work in several cases. 

First, let us choose two particular sectors of eigenvalues, fixing $n$ and $m$. It is often possible to bound the eigenvalue differences for
this sector as
\begin{eqnarray}
   |\lambda_{k,n}-\lambda_{j,m}| \geq c_{n,m} \left| \cos p_j - \cos p_k  \right|
   = 2c_{n,m}  \left| \sin \left(\frac{\pi(k-j)}{K}\right) \right| \left| \sin \left(\frac{\pi(k+j)}{K}\right) \right|,
\label{cosbound}
\end{eqnarray}
for some constant $c_{n,m}$, where 
where $p_k=\frac{2\pi k}{K}$ are the momenta, and the indices $j,k$ run from $0$ to $K-1$, observing $|k-j|\neq
\{0,\frac{K}{2}\}$. 
We will rewrite \eqref{cosbound} using the substitution $a=j+k$ and $b=k-j$. 
\begin{eqnarray}
    \sum_{\lambda_{j,m}\neq\lambda_{k,n}} 
    \frac{1}{|\lambda_{j,m}-\lambda_{k,n}|}
    \leq 
   \frac{1}{2c_{n,m}}
   \sum_{b}
   \sum_{a}
      \frac{1}{
       \left| \sin \frac{\pi b}{K} \right|
       \left| \sin \frac{\pi a}{K} \right| }
\end{eqnarray}
where $|b|\leq K-1$ while $b\neq 0$, and $|b|\leq a \leq 2(K-1)-|b|$ while $a\not\in \{0,K\}$.
First, because of symmetry, we observe that it is enough to sum only over $0<b\leq \frac{K}{2}$ and multiply 
the resulting sum by 4. Second, it can only increase our upper bound if we count all $a>0$, instead of having to take care with counting starting
at $|b|$. The symmetry of the term involving $a$ then again allows us to sum only up to $a=\frac{K}{2}$ and multiply the result by 4.
Therefore, we obtain
\begin{eqnarray}
    \sum_{\lambda_{j,m}\neq\lambda_{k,n}} 
    \frac{1}{|\lambda_{j,m}-\lambda_{k,n}|}
    \leq 
   \frac{1}{16c_{n,m}}
   \left(
   \sum_{b=1}^{K/2}
      \frac{1}{
       \left| \sin \frac{\pi b}{K} \right|}\right)
   \left(\sum_{a=1}^{K/2}
      \frac{1}{
       \left| \sin \frac{\pi a}{K} \right| }\right).
\end{eqnarray}
Recalling now that $\sin b \geq \frac{2b}{\pi}$, we can deal with each sum as
\begin{eqnarray}
   \sum_{b=1}^{K/2}
      \frac{1}{\left|\sin\frac{\pi b}{K}\right|}
   \leq
   \frac{K}{2}
   \sum_{b=1}^{K/2}
      \frac{1}{b}
   \leq
   \frac{K}{2} \ln \frac{K}{2},
\end{eqnarray}
thus finally expressing the sum in \eqref{Lemma43} (note that we worked only in a single $n,m$ sector) as
\begin{eqnarray}
  \sum_{\lambda_{k,n} \neq \lambda_{j,m}}
      \frac{2
         \left| 
         \braket{\psi_k}{\varphi_0}
         \right|^2
      }{T|\lambda_{k,n}-\lambda_{j,m}|}
      \leq
      \frac{2}{TK} \frac{K^2\ln^2\frac{K}{2}}{16 c_{n,m}}
      =\frac{1}{8 c_{n,m}} \frac{K}{T} \ln^2\frac{K}{2},
\end{eqnarray}
where the extra factor $K^{-1}$ comes from the term $\left| \braket{\psi_k}{\varphi_0}\right|^2$, when we expect the initial state to have roughly
equal overlap with all momentum states.
According to \eqref{Lemma43} and working this out for all sectors $n,m$, this suffices to show an upper bound on the mixing time (in the
time-averaged sense) for this type of quantum walk, which grows with the system size a $T_{mix}(\ep) \leq O(\ep^{-1} K \ln^2 {K})$.

We will now show that for particular examples of walks on necklaces, the eigenvalues
obey \eqref{cosbound}, and so that we can use the above approach for proving their convergence.
The first example that we can deal with using this method is the cycle itself, where \eqref{cosbound} is an equality. Therefore, we have just shown
that the time-averaged distribution (when starting from a single point) converges to the limiting distribution for a cycle of length $K$ with
mixing time $T_{mix}(\ep) \leq O^*\left(\ep^{-1} K\right)$ (up to logarithmic factors).


\section{Examples: Quantum Walks on Comb-like Necklaces}
\label{sec:examples}

\begin{figure}
   \begin{center}
      \includegraphics[width=2.5in]{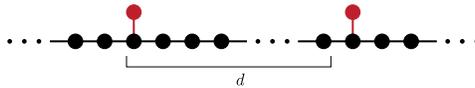}
   \end{center}
   \caption{A $(K,d)$-comb graph is a ring of length $Kd$, with an extra vertex connected at every $d$-th node.}
   \label{figurecombKd}
\end{figure}

We now look at a specific type of necklaces, which appear in the quantum-walk based model of computation \cite{railroadswitch2}.
These ``combs'' are constructed from a ring of length $K d$ by attaching an extra node (tooth) to the ring at every $d$-th vertex as in Fig.~
\ref{figurecombKd}. The ``pearl'' in this comb-like graph has size $d+1$, and there are $K$ of them, so 
the total number of vertices in this graph is $N = K(d+1)$. We will now analyze the spectra 
and mixing properties on the $(K,d)$-comb necklaces, showing their similarity to (and differences from) a simple cycle.


\subsection{The $(K,1)$-comb necklace}
\label{sec:K1}
This is the simplest of the graphs, with a pearl that has only two nodes (the base and the tooth):
\begin{eqnarray}
   P^{(K,1)} = 
   \left[
      \begin{array}{cc}
         0 & 1  \\
         1 & 0 
      \end{array}
   \right].
\end{eqnarray}
Because $P$ has a single root, the matrix $Q_k$ needed to construct $Y_k$ \eqref{whattodiag} reads
\begin{eqnarray}
   Q^{(K,1)}_k = 
   \left[
      \begin{array}{cc}
         e^{ip_k} + e^{-ip_k} & 0 \\
         0 & 0 
      \end{array}
   \right].
\end{eqnarray}
Therefore, the matrix $Y_k$ \eqref{whattodiag} is
\begin{eqnarray}
   Y^{(K,1)}_k = 
   \left[
      \begin{array}{cc}
         e^{ip_k} + e^{-ip_k} & 1 \\
         1 & 0 
      \end{array}
   \right].
\end{eqnarray}
Its eigenvalues are
\begin{eqnarray}
   \lambda^{(K,1)}_{k,\pm} = \cos p_k \pm \sqrt{1+\cos^2 p_k},
   \label{combK1}
\end{eqnarray}
and the corresponding eigenvectors are
\begin{eqnarray}
   \kets{y^{\,(k,\pm)}} = \frac{1}{\sqrt{1+\lambda_{k,\pm}^2}} \left[ 
      \begin{array}{c}
      \lambda_{k,\pm} \\
      1 
      \end{array}
   \right]
\end{eqnarray}
where the upper vector component corresponds to the base (and the lower component to the tooth) of the comb.
According to \eqref{ceigenfunctions}, this gives us the eigenvectors of $H$ as 
\begin{eqnarray}
   \kets{\psi^{(k,\pm)}} = 
   \frac{1}{\sqrt{K}} \sum_{j=1}^{K}  
   \frac{e^{ip_k j}}{\sqrt{1+\lambda_{k,\pm}^2}}
   \left[
   \begin{array}{c}
      \lambda_{k,\pm}\\
      1
      \end{array}
   \right]_{(j)}
\end{eqnarray}
with $p_k = \frac{2\pi k}{K}$ for $k=0,\dots,K-1$. 
The eigenvalues $\lambda_k$ \eqref{combK1} of the Hamiltonian are symmetrically distributed around zero, and each of
them is also doubled if $K$ is even. Note that two of the eigenvectors $\ket{\psi}$ are zero on every other base and
tooth, and correspond to eigenvalues $\pm 1$.

The limiting distribution is analyzed in Appendix \ref{app:comb1}. We find that for large $K$, the 
limiting distribution (when starting from a base vertex) is  $\frac{4-\sqrt{2}}{4K}$ on base vertices and $\frac{\sqrt{2}}{4K}$ on
teeth, with corrections for the initial vertex and the vertex across from it.
We  now prove convergence to the time-averaged limiting distribution, showing that the (time-averaged) mixing on this densest comb is no different
than the one we saw for a cycle. We will upper bound the sum in \eqref{Lemma43} by the method in Section \ref{sec:method},
dividing the eigenvalues into 4 regions, $++, +-, -+, --$, corresponding to choices of $n$ and $m$.
In the $+-$ and $-+$ regions, we have
$|\lambda_{j,0}-\lambda_{k,1}| \geq const.,$
so the inverse of such terms does not govern the scaling in \eqref{Lemma43}. 
The important region combinations must then be $++$ and $--$, where a few lines of algebra give us
\begin{eqnarray}
      \left| \lambda_{j,\pm}-\lambda_{k,\pm} \right| &=&  
             \frac{\left| \cos{p_j}- \cos{p_k}\right| 
             .
       \left| \lambda_{j,\pm} + \lambda_{k,\pm}\right|}{\sqrt{1+\cos^2 p_j} + \sqrt{1+\cos^2 p_k}} 
                   \geq  \frac{\sqrt{2}-1}{\sqrt{2}} \left| \cos{p_j}-\cos{p_k} \right|,
\end{eqnarray}
as the eigenvalues are well bounded away from zero. Armed with this inequality, 
and the fact that the overlap of a single-starting-vertex initial state with the eigenvectors scales as $\frac{1}{\sqrt{K}}$,
we can now use the result of Section \ref{sec:method}. 
This gives us an upper bound on the mixing time for the $(K,1)$-comb necklace, scaling with $K$ as $T^{(K,1)}_{mix}(\ep) \leq
O^{*}\left( \ep^{-1} K
\right)$, i.e. the same as for a cycle with no teeth, up to logarithmic factors.


\subsection{The $(K,2)$-comb}
The next example is the $(K,2)$ comb. It has a tooth (extra vertex) at every second node of the basic loop, so its pearl
$P$ has three vertices. We label the base of the tooth as vertex 1, and the top of the tooth as vertex 2, giving: 
\begin{eqnarray}
   P = 
   \left[
      \begin{array}{ccc}
         0 & 1 & 1 \\
         1 & 0 & 0 \\
         1 & 0 & 0
      \end{array}
   \right].
\end{eqnarray}
Following the procedure of Section \ref{sec:ansatz},
we need to find the eigenvalues and eigenvectors of the matrices $Y^{(K,2)}_k$ constructed as in \eqref{whattodiag}:
\begin{eqnarray}
   Y^{(K,2)}_k = 
   \left[
      \begin{array}{ccc}
         0 & 1 & 1 \\
         1 & 0 & 0 \\
         1 & 0 & 0
      \end{array}
   \right]+
   \left[
      \begin{array}{ccc}
         0 & 0 & e^{-ip_k} \\
         0 & 0 & 0 \\
         e^{ip_k} & 0 & 0
      \end{array}
   \right]
=
   \left[
      \begin{array}{ccc}
         0 & 1 & e^{-\frac{ip_k}{2}} 2\cos \frac{p_k}{2} \\
         1 & 0 & 0 \\
         e^{\frac{ip_k}{2}} 2\frac{\cos p_k}{2} & 0 & 0
      \end{array}
   \right]
   .
\end{eqnarray}
After some algebra, we find that its three eigenvalues are
\begin{eqnarray}
   \lambda^{(K,2)}_{k,0} = 0, \qquad 
   \lambda^{(K,2)}_{k,\pm}=\pm \sqrt{3+2\cos p_k},
\end{eqnarray}
with the corresponding eigenvectors
\begin{eqnarray}
   \kets{y^{\,(k,0)}} = \frac{1}{\sqrt{(3+2\cos p_k)}} \left[ 
    \begin{array}{c}
        0 \\
        2 e^{-\frac{ip_k}{2}}\cos \frac{p_k}{2} \\
        -1\\
      \end{array}
   \right],
   \qquad
      \kets{y^{\,(k,\pm)}} = \frac{1}{\sqrt{2 (3+2\cos p_k)}} \left[ 
      \begin{array}{c}
      \pm \sqrt{3+2\cos p_k} \\
       1 \\
       2 e^{-\frac{ip_k}{2}} \cos\frac{p_k}{2} \\
      \end{array}
   \right]. 
   \label{K2eigenvectors}
\end{eqnarray}
To construct the eigenvectors of the Hamiltonian $H$, we use \eqref{K2eigenvectors} in equation \eqref{ceigenfunctions}.

Let us now look for a lower bound on the gap between eigenvalues. 
When we choose two eigenvalues from different sectors ($0$, $+$ or $-$), the differences between them are
always larger than $1$. The only interesting cases are thus the $++$ and $--$ choices of eigenvalue pairs. There we find
\begin{eqnarray}
      \left| \lambda_{j,\pm}-\lambda_{k,\pm} \right|
      = \left| \sqrt{3+2\cos{p_k}}-\sqrt{3+2\cos{p_j}}\right|
      = \frac{2\left(\cos{p_k}-\cos{p_j}\right)}{\sqrt{3+2\cos{p_k}}+\sqrt{3+2\cos{p_j}}}\geq \frac{1}{2}\left( \cos{p_k}-\cos{p_j} \right).
\end{eqnarray}
This lower bound on the nonzero eigenvalue gaps allows us to use the results of Section \ref{sec:method} and prove the upper bound 
$T^{(K,2)}_{mix}(\ep) \leq O^*\left( \ep^{-1} K \right)$
on the mixing time for the $(K,2)$-comb. This is once again the same upper bound we found for the cycle and the $(K,1)$-comb in Section \ref{sec:K1}.


\subsection{The $(K,d)$-combs}

\begin{figure}
   \begin{center}
      \includegraphics[width=3in]{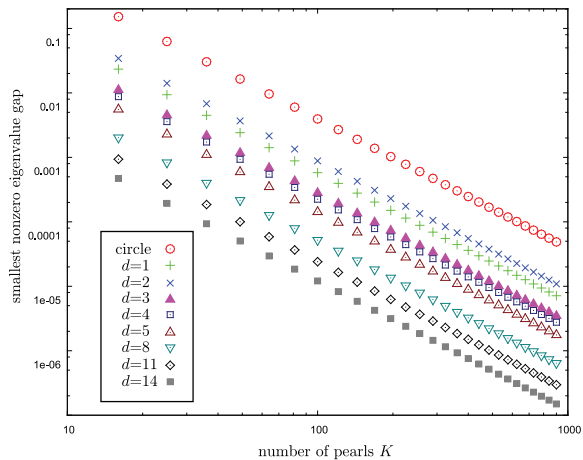}
   \end{center}
   \caption{A log-log plot of the smallest nonzero eigenvalue differences for $(K,d)$-comb graphs (see Figure \ref{figurecombKd}) and a cycle. A straight-line fit through the datapoins indicates a $K^{-2}$ scaling with the growing number of pearls.}
   \label{figurecombeigs}
\end{figure}

In the last example we want to show that comb-like necklaces with $d$ vertices between teeth (see Figure \ref{figurecombKd}) mix similarly to cycles.
We dealt with the most non-cycle-like examples in the previous Sections, and now we will numerically look at combs with general spacing $d$. The
results for the smallest nonzero eigenvalue differences for $d\leq 15$ combs are shown in Figure \ref{figurecombeigs}. In a log-log plot of
the smallest eigenvalue difference vs. the number of pearls $K$ (for various values of $d$), we see the characteristic $K^{-2}$ scaling.
Thus, the numerics imply that the scaling of the mixing time gets no worse than 
$T^{(K,d)}_{mix}(\ep) = O\left( \ep^{-1} K^2 \log^2K \right)$. However, it is likely that the eigenvalue
differences also obey the cos-like scaling \eqref{cosbound} as we have seen for $d=1,2$. If we could show this, we would again prove that the mixing time scales with $K$ as $T^{(K,d)}_{mix}(\ep) = O^*\left( \ep^{-1}  K \right)$.

\section{Conclusions}
\label{sec:conclusions}

The goal of this paper was to utilize the cyclical repetitive structure of necklace-like graphs, providing a general method for analyzing the eigenvectors and eigenvalues of continuous-time quantum walks on such graphs. Using a Bloch-theorem-like ansatz, we block-diagonalized the Hamiltonian, decreasing the effective size of the problem from $KM$ to $M$, where $M$ is the size of a pearl and $K$ is the number of pearls in the necklace. Next, we wanted to investigate the mixing times (for approaching a limiting distribution in a time-averaged sense) for these quantum walks. In Section \ref{sec:method} we have shown that proving a $\cos$-like lower bound \eqref{cosbound} on (non-equal) eigenvalue differences results in a mixing time 
$T_{mix}(\ep) \leq O^*\left(\ep^{-1} K\right)$ for a graph with $K$ pearls, which is the same as for a cycle with $K$ nodes. Note though, that the prefactor in the mixing time can depend on the size of the pearl $M$. Finally, in Section \ref{sec:examples} we exhibited the bound \eqref{cosbound} (and thus the cycle-like mixing time) for two necklace-like graphs. These graphs appear in the models of quantum computation \cite{railroadswitch, railroadswitch2} that extend the Feynman-computer with the so-called railroad switches, and finding the polynomial-time (in $K$) scaling of the mixing-time is required for showing their effectiveness. 


\section{Acknowledgements}
\label{sec:thanks}

DN acknowledges support from the Slovak
Research and Development Agency under the contract No. LPP-0430-09,
from the projects meta-QUTE ITMS
26240120022, VEGA QWAEN and European project
Q-ESSENCE.
We thank Zuzana Gavorov\'a, Daniel Reitzner, and Vladim\'ir Bu\v{z}ek for fruitful discussions.



\appendix
\section{The limiting distribution for a $(K,1)$-comb necklace}
\label{app:comb1}

\begin{figure}
   \begin{center}
        \includegraphics[width=2.8in]{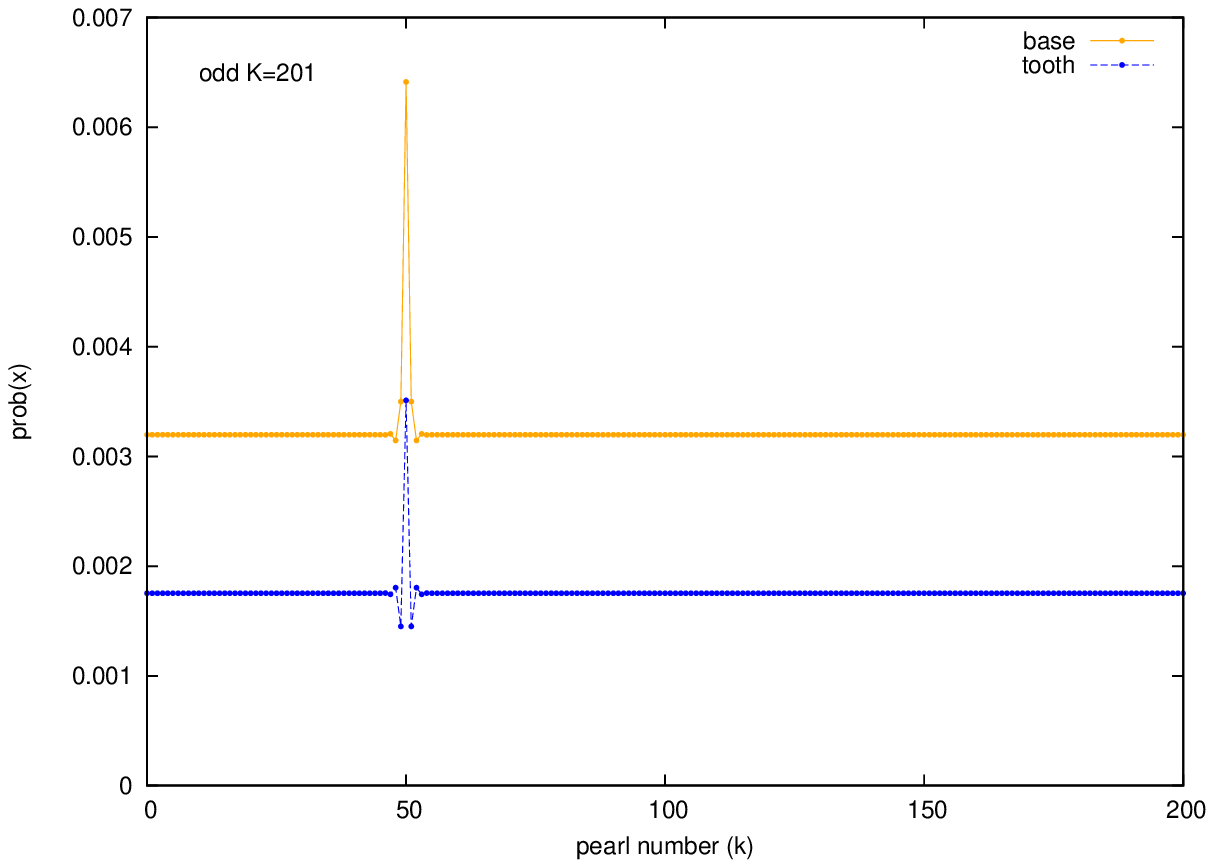}
        \includegraphics[width=2.8in]{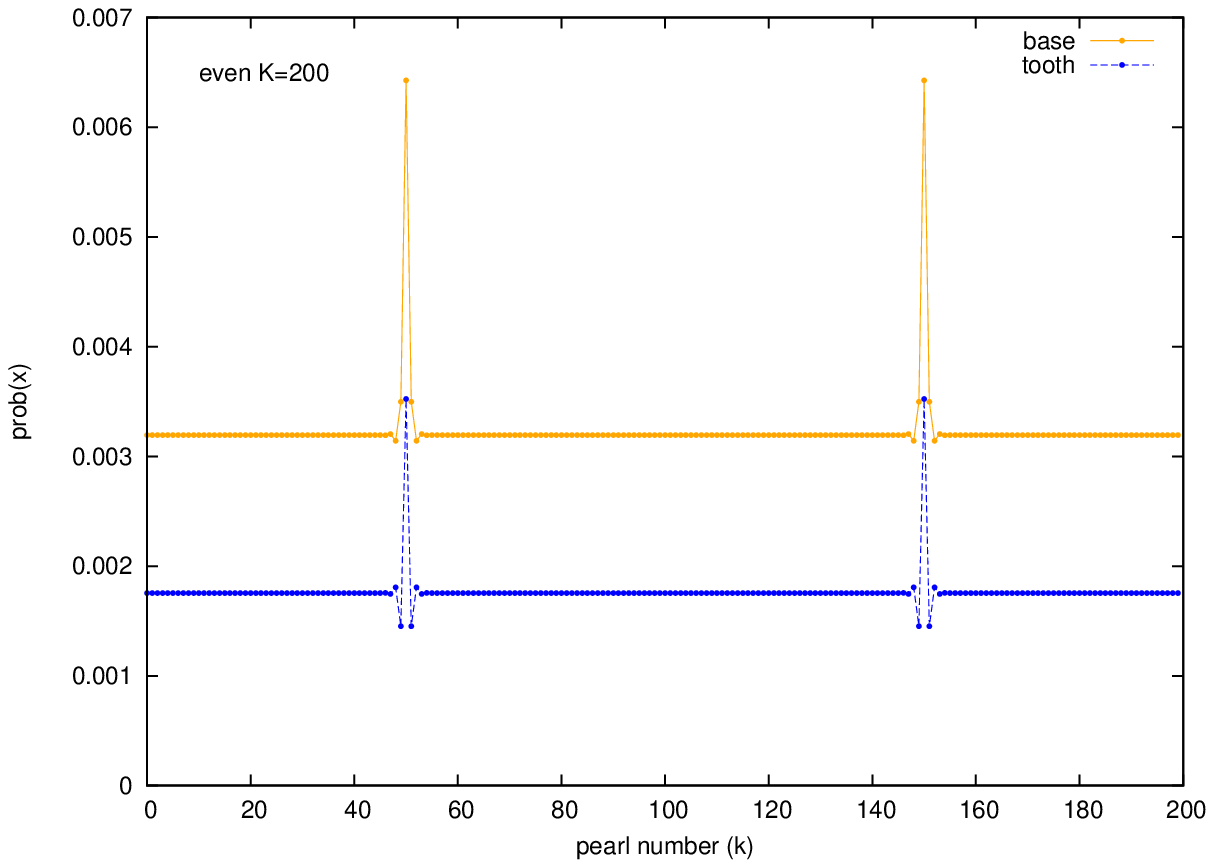}
   \end{center}
   \caption{The limiting distribution \eqref{limprob} on bases and teeth of a $(K,1)$-comb when starting from the base of the $50-$th pearl for odd $K=201$
(left) and even $K=200$ (right).}
   \label{K1lim}
\end{figure}

We now analyze the limiting distribution for the continuous-time quantum walk on a $(K,1)$-comb in Fig.~\ref{K1lim},
starting from a single (base) vertex $z$. First, we do our work analytically, and end with a few high-$K$ numerical approximations. 
The conclusion is that the distribution is flat except for a few points very close to the two special positions $x^* = z$ and $|x^*-z|=\frac{K}{2}$.

First, we look at the time-averaged limiting distribution for going from base $x_b$ to base $z_b$ (which thanks to the identity \eqref{l1} turns out to be the same as for going
from
tooth $x_t$ to tooth
$z_t$), obtaining
\begin{equation}
\begin{array}{ll}
   \pi\left(x_b|z_b\right) = \pi\left(x_t|z_t\right) 
   &= \sum_{k,n} \psi_{x_b}^{(k,n)} \psi_{y_b}^{(k,n)*} \psi_{x_b}^{(k,n)*} \psi_{y_b}^{(k,n)}
      + \sum_{k,n} \psi_{x_b}^{(k,n)} \psi_{y_b}^{(k,n)*} \psi_{x_b}^{(K-k,n)*} \psi_{y_b}^{(K-k,n)} \\
    \noalign{\medskip}
      &- \sum_{n} \psi_{x_b}^{(0,n)} \psi_{y_b}^{(0,n)*} \psi_{x_b}^{(0,n)*} \psi_{y_b}^{(0,n)} 
      - \sum_{n} \psi_{x_b}^{(K/2,n)} \psi_{y_b}^{(K/2,n)*} \psi_{x_b}^{(K/2,n)*} \psi_{y_b}^{(K/2,n)},
\end{array}
\label{eqA1}
\end{equation}
where the last term involving $K/2$ occurs only for even $K$. Using the identities
\begin{eqnarray}
&&\lambda_{k,+} \lambda_{k,-} = -1\\ \label{lambdaid}
&&\sum_{n\in\{+,-\}} \frac{1}{\left(1+\lambda_{k,n}^2\right)^2} = 1 - \frac{1}{2\left(1+\cos^2 p_k\right)} 
   = \sum_{n\in\{+,-\}} \frac{\lambda_{k,n}^4}{\left(1+\lambda_{k,n}^2\right)^2} \label{l1}
\end{eqnarray}
and denoting
\begin{eqnarray}
A &=& \frac{1}{K}\sum_{k=0}^{K-1} \frac{1}{2\left(1+\cos^2 p_k\right)}, \\
B_{(x|z)} &=& \frac{1}{K}\sum_{k=0}^{K-1} \frac{1}{2\left(1+\cos^2 p_k\right)} e^{\frac{i
2\pi}{K}2(x-z)k},
\end{eqnarray}
\begin{equation}
  C = \left\{
  \begin{array}{l l}
     &\frac{3}{4K}\quad \text{for odd $K$ },\\
     \noalign{\medskip}
     &\frac{3}{2K} \quad \text{for even $K$},
  \end{array} \right.
\end{equation}
with $f_{x,z}$ defined in \eqref{fxz},
we rewrite \eqref{eqA1} to finally obtain
\begin{equation}
   \pi\left(x_b|z_b\right) = \pi\left(x_t|z_t\right) 
   = \frac{1}{K}\left(1-A-B_{(x|z)}+C+f_{x,z}\right).
\end{equation}

Next, we compute the time-averaged limiting distribution for the ``start from a base -- go to a tooth'' transition, using the identity  
\begin{eqnarray}
 L_k^{(2)} = \sum_{n\in\{+,-\}} \frac{\lambda_{k,n}^2}{\left(1+\lambda_{k,n}^2\right)^2} &=& \frac{1}{2\left(1+\cos^2
p_k\right)}.
\end{eqnarray}
and \eqref{lambdaid}.
We obtain
\begin{equation}
    \pi\left(x_b|y_t\right) =\pi\left(x_t|y_b\right)=
    \left\{ \begin{array}{l l}
        &\frac{1}{K}\left(A+B_{(x|z)}+\frac{1}{4K}\right)   \quad \text{for odd $K$},  \\
        \noalign{\medskip}
        &\frac{1}{K}\left(A+B_{(x|z)}+\frac{1}{2K}\right)  \quad \text{for even $K$}.  
\end{array}\right.
\end{equation}

Finally, let us look at the high $K$ approximation (see also Fig.\ref{K1lim}). In $A$  we replaced the sum by an integral and obtained
$A\approx\frac{\sqrt{2}}{4}$. At points $z$ for which $(x-z)\in\{0,\frac L2\}$, the expression $B_{(x|z)}$ equals $A$ exactly, while it rapidly falls off to zero with
growing distance of $z$ from $x$ (or $\frac{K}{2}+x$). The limiting distribution when starting from a base point for even number of pearls is thus
well approximated by a flat distribution with $\frac{4-\sqrt{2}}{4K}  -\frac{6}{4K^2}$ on
bases and $\frac{\sqrt{2}}{4K} - \frac{1}{2K^2}$ on
teeth, with the exception of the starting pearl and the pearl exactly opposite to it receiving 
$\frac{4-\sqrt{2}}{2K} -\frac{6}{4K^2}$ and
$\frac{\sqrt{2}K}{2K} -\frac{1}{K^2}$, respectively. It is very similar for odd $K$, except that we do not have the special case of the opposite pearl. For an example of the limiting distribution with odd and even $K$, see Fig.~\ref{K1lim}.


\begin{thebibliography}{99}


\bibitem{metropolis}
N.~Metropolis, A.W.~Rosenbluth, N.N.~Rosenbluth, A.H.~Teller, E.~Teller: \textsl{Equation of State Calculations for Fast Computing Machines,}
J.~Chem.~Phys.~\textbf{21,} 1087 (1953)

\bibitem{metropolis2}
W.K.~Hastings: \textsl{Monte Carlo Sampling Methods Using Markov Chains and Their Applications,} Biometrika \textbf{57,} 97 (1970)

\bibitem{annealing}
S.~Kirkpatrick, C.D.~Gelatt Jr., M.P.~Vecchi: \textsl{Optimization by Simulated Annealing,} Science \textbf{220,} 671 (1983)

\bibitem{annealing2}
J.~\v Cern\'y: \textsl{Thermodynamical Approach to the Travelling Salesman Problem,} J.~Opt.~Theory Appl.~\textbf{45,} 41 (1985)

\bibitem{Jerrum2}
		M.~Jerrum and A.~Sinclair, {\em Polynomial-Time Approximation
		Algorithms for the Ising Model}, SIAM Journal on Computing, vol.~22,
		pp.~1087--1116, 1993.

\bibitem{Vigoda}
		M.~Jerrum, A.~Sinclair, and E.~Vigoda, {\em A Polynomial-Time
		Approximation Algorithm for the Permanent of a Matrix Non-Negative
		Entries}, Journal of the ACM, vol.~51, issue 4, pp.~671--697, 2004.

\bibitem{AAKVwalk}
D.~Aharonov, A.~Ambainis, J.~Kempe, U.~Vazirani: \textsl{Quantum walks on graphs,} in STOC '01: Proceedings of the thirty-third annual ACM symposium on Theory of computing, 50--59 (New York, NY, USA, ACM, 2001)

\bibitem{FarhiWalk}
E. Farhi, S. Gutmann, {\sl Quantum Computation and Decision Trees}, Phys. Rev. A 58 (2), 915 (1998)

\bibitem{ABNVWwalk}
A.~Ambainis, E.~Bach, A.~Nayak, A.~Vishwanath, and J.~Watrous: \textsl{One-dimensional quantum walks,} in STOC '01: Proceedings of the thirty-third annual ACM symposium on Theory of computing, 37--49 (New York, NY, USA, ACM, 2001)

\bibitem{Santha}
		M.~Santha, {\em Quantum Walk Based Search Algorithms}, Proc. of 5th Theory and
		Applications of Models of Computation (TAMC08), Lectures Notes on Computer
		Science, vol.~4978, pp.~31--46, 2008.

\bibitem{gluedtrees}  
A. M. Childs, R. Cleve, E. Deotto, E. Farhi, S. Gutmann, D. A. Spielman, {\sl Exponential algorithmic speedup by quantum walk},
Proceedings of the 35th ACM Symposium on Theory of Computing, pp. 59-68 (2003)

\bibitem{nandtrees}
E. Farhi, J. Goldstone, S. Gutmann, {\sl A quantum algorithm for the Hamiltonian NAND tree}, Theory of Computing, Vol. 4, no. 1, pp.169-190, 2007;
quant-ph/0702144

\bibitem{universal} 
A. M. Childs, {\sl Universal computation by quantum walk} 
Physical Review Letters 102, 180501 (2009)

\bibitem{2Dmix}
F. L. Marquezino, R. Portugal, Mixing times in quantum walks on two-dimensional grids, 
Physical Review A 82, 042341 (2010)

\bibitem{hypercubemix}
C. Moore, A. Russell, Proc. of the 6th International Workshop on Randomization and Approximation Techniques (RANDOM 2002), Cambridge, MA, LNCS Vol. 2483, Springer-Verlag, Berlin, 2002, pp. 164-178

\bibitem{circulant1}
A. Ahmadi, R. Belk, C. Tamon, C. Wendler, On mixing in continuous-time quantum walks on some circulant graphs, 
Quantum Information \& Computation, Vol.3, No.6 (2003), 611-618

\bibitem{circulant2}
P. Lo, S. Rajaram, D. Schepens, D. Sullivan, C. Tamon, J. Ward, Mixing of Quantum Walk on Circulant Bunkbeds, Quantum Information and Computation, Vol. 6, No. 4\&5 (2006), 370-381

\bibitem{HamiltonComplex}
T. Osborne, Hamiltonian complexity, arXiv:1106.5875 (2011)


\bibitem{NCbook}
M. A. Nielsen, I. L. Chuang, Quantum Information and Computation, Cambridge University Press, Cambridge, UK, 2000.

\bibitem{Feynman85}
R.~Feynman, Quantum mechanical computers, Opt. News, vol. 11, pp.~11--46 (1985),

\bibitem{KKR}
J.~Kempe, A.~Kitaev, and O.~Regev.
\newblock The complexity of the local {H}amiltonian problem.
\newblock { SIAM Journal of Computing}, 35(5):1070--1097, 2006.


\bibitem{new3local}
D.~{Nagaj} and S.~{Mozes},
{\em New construction for a QMA complete three-local Hamiltonian}, Journal of Mathematical Physics, 48:2104, 2007.
   
\bibitem{deFalco}
D. de Falco, D. Tamascelli, {\em Grover's algorithm on a Feynman computer}, 
J. Phys. A: Math. Gen. 37, 909-930 (2004)

\bibitem{railroadswitch}
D.~Nagaj, {\em Fast Universal Quantum Computation with Railroad-switch Local Hamiltonians}, Journal of Mathematical Physics, 51 (6), 062201 (2010)

\bibitem{LidarMizel}
A.~Mizel, D.~A.~Lidar and M.~Mitchell, Simple Proof of Equivalence Between Adiabatic Quantum Computation and the Circuit Model, Phys. Rev. Lett.
99, 070502 (2008).

\bibitem{qma1d}
D. Aharonov, D. Gottesman, S. Irani, J. Kempe, The power of quantum systems on a line, Proc. 48th IEEE FOCS, 373-383 (2007)

\bibitem{railroadswitch2}
D.~Nagaj, {\em Universal 2-local Hamiltonian Quantum Computing}, Phys. Rev. A 85, 032330 (2012)

\bibitem{EldarRegev}
L.~Eldar, O.~Regev, Quantum SAT for a Qutrit-Cinquit Pair is QMA$_1$-Complete, ICALP 2008, L. Aceto et al. (Eds), Part I, LNCS 5125, pp. 881-892, Springer-Verlag Berlin, Heidelberg (2008),
L.~Eldar, P.~Love, D.~Nagaj, O.~Regev, in preparation

\bibitem{HQCA1d}  
D.~Nagaj and P.~Wocjan, Hamiltonian Quantum Cellular Automata in 1D, Phys. Rev. A 78, 032311 (2008)

\end{thebibliography}
\end{document}